\documentclass[12pt,onepage,oneside]{article}
\usepackage{fullpage}
\usepackage{amsfonts,amsmath,amssymb,eqnarray}
\usepackage{graphicx}
\usepackage{dcolumn}
\usepackage{siunitx}
\usepackage{tabularx}
\usepackage{booktabs}
\usepackage{xcolor,colortbl}
\usepackage{color}
\usepackage{comment}
\usepackage{setspace}
\usepackage{bbm}
\usepackage{natbib}
\usepackage{threeparttable}
\usepackage{multirow}
\usepackage{lscape}
\usepackage{pdflscape}
\usepackage{rotating}
\usepackage{graphicx}
\usepackage{times}
\usepackage{float}
\usepackage[titletoc,toc,title]{appendix}
\usepackage{subcaption}
\usepackage{color}
\usepackage{makecell}

\usepackage{amsopn}
\usepackage[colorlinks=true,linkcolor=black,allcolors=black]{hyperref}
\graphicspath{ {LateX} }
\usepackage{pdflscape}
\usepackage{float}
\usepackage[titletoc,toc,title]{appendix}
\usepackage{amsmath}
\usepackage{authblk}
\graphicspath{ {LateX} }
\usepackage[left=1in,right=1in,top=1.in,bottom=1.in]{geometry}
\usepackage{times}

\begin{document}

\title{Investment Disputes and Abnormal Volatility of Stocks\footnote{Authors gratefully acknowledge support from the Czech Science Foundation under the GA 18-04630S project. Authors order is alphabetical.}}

\author[2]{Jozef Barun\'{\i}k\thanks{barunik@fsv.cuni.cz}}
\author[1]{Zden\v{e}k Dr\'{a}bek}
\author[2]{Mat\v{e}j Nevrla}
\affil[1]{CERGE-EI}
\affil[2]{Charles University and Czech Academy of Sciences}

\date{}

\maketitle
\begin{abstract}
\noindent Dramatic growth of investment disputes between foreign investors and host states rises serious questions about the impact of those disputes on investors. This paper is the first to explain increased uncertainty of investors about the outcome of arbitration, which may or may not lead to compensation for damages claimed by the investor. We find robust evidence that investment disputes lead to abnormal share fluctuations of companies involved in disputes with host countries. Importantly, while a positive outcome for an investor decreases uncertainty back to original levels, we document strong increase in the volatility of companies with negative outcome for the investor. We find that several variables including size of the award, political instability, location of arbitration, country of origin of investor or public policy considerations in host country explain large portion of the investor's uncertainty.
\end{abstract}

\section{Introduction}

Following the rapid growth of foreign direct investment (FDI) over the last 30 years or so, investment disputes between investors and host country states have also increased dramatically. This is serious since FDI have been the key factor of global economic growth during the period, and investment disputes can jeopardize economic prosperity, global fight against poverty, diseases and other public mal \citep{gonzalezkusek2018,newman2015technology,baldwin2015supply}. Investment disputes have many origins ranging from direct expropriation of investors assets to disputes about various regulatory interventions impeding business of the investors. They all can seriously affect the profitability of FDI. The resolution of disputes takes place under the legal provisions of different international investment agreements (IIA) and, most frequently, under bilateral investment agreements (BIT), which stipulate, \textit{inter alia} the rules and procedures for the resolution of disputes through international arbitration. The main purpose of these agreements is to protect foreign investors even though some, especially some of the more recent ones include also provisions to improve market access for foreign firms \citep{dolzer2012principles,juillard2000freedom}.\footnote{The level of protection also varies among BITs \citep{alschner2016mapping}. Note that the breach of contract can ``normally'' only be initiated by investors and host countries typically do not have the right to challenge breaches by investors. The latter are subject to standard commercial courts.}

Investment disputes can be very costly \citep{hajzler2012expropriation}. Host countries may suffer adverse reputational costs, high litigation costs and high costs of compensations. For many poor countries, the costs of litigation can be excessive and the arbitration is sometimes accessible only with the help of foreign aid.  \textit{Pari passu}, an involvement of investors in disputes with foreign governments can increase uncertainties about the conduct of their business, and the market may perceive the firm as a greater risk. Should investors lose the arbitration, the disputes will lead not only to litigation costs but also to lost income. The relative costs to investors will depend on the value of the claim relative to the size of the balance sheet of the firm and the actual outcome of the dispute. 

The empirical literature dealing with investment disputes is extremely limited.  It has been almost exclusively concerned with the analysis of the effects of IIA on FDI. Most of the research is not strictly about investment disputes because IIA are not only about investment protection but also about market access, as noted above. Economic studies of investment disputes are few and all are concentrated on three specific issues. First, the research has been entirely concentrated on host countries and their benefits from FDI.\footnote{The literature has been reviewed, for example, in \cite{pohl2018societal} and the empirical evidence is synthesized in \cite{brada2020two}.} The second important stream of the empirical literature has involved an assessment of the performance of international arbitrations \citep{franck2007empirically,van2011use,schultz2014investment}. The literature provides an empirical assessment of the litigation explosion, focusing on (1) identifying the parties to disputes, and the subject of the arbitration; (2) increases in awards, (3) analysis of win/loss rates, (4) of amounts claimed and awarded, (5) arbitration costs. (6) Attempts were also made to compare different dispute resolution processes, (7) the role of nationality and gender of arbitrators  and (8) assess the fairness and independence in investment arbitration. They also ask about the costs of arbitrations – whether the costs are excessive in relation to benefits and how they relate to the levels of income. Third, an alternative approach to the study of investment disputes is a recent attempt to link macroeconomic indicators and disputes \citep{bellak2020}.\footnote{\cite{bellak2020} also review the relevant literature. }

On the other hand, there is no empirical literature that would examine the impact of FDI disputes on investors.\footnote{The only exception is a very recent attempt of \cite{brada2020one} who look in their event study at the effects of arbitral decisions on the real returns of investing companies but the study has not yet been published. The study currently exists as a working paper and is under review for publication. The authors find that ``in cases where the arbitrators rule in favor of the MNC and award it damages, the stock price and the value of the firm increase by 3 percent over the ten-day period following the announcement of the arbitrators decision.... In cases, where the arbitration tribunal ruled against the MNC and awarded no monetary damages, the stock price of the firm declined by 2 percent''.} This is unfortunate because the disputes can affect the value of firms as well as their incentives to relocate abroad. In addition, as disputes are very likely to affect the shares of investor's firm, they will also impact their trading in the market. We intend to explore this new area of research. 

The aim of our study is to assess the impact of investment disputes on the volatility of the company's shares. The risks and uncertainties originating in investment disputes are likely to be reflected in changes in the prices of shares of investors in the market. One approach to the evaluation of the effects of disputes on investors is to estimate real rates of return of the investors. Our approach is different. We document the impact of an investment dispute on the uncertainty and on the stock price measured by the second moment, its volatility. As uncertainty rises about business operations and hence about the risk of the stock, stock price volatility is likely to increase with the emergence of disputes. Positive resolution of disputes in favor of the investor can be expected to lead to a reduced uncertainty and, consequently, to lower stock volatility. 

Academics, policy makers, traders and other financial practitioners have been trying to understand the uncertainty regarding future price fluctuations measured by volatility for decades.\footnote{For a comprehensive early survey of the literature see \cite{shiller1992market}.} Whereas origins of price movements may have good economic sense or can be explained by economic fundamentals or behavioral psychology, its ultimate causes are poorly known in all speculative markets. Rapidly changing fluctuations can hardly be attributed to a single cause.  The economic sources of movements typically include fundamental shocks affecting the economy, the shocks to technology, to consumer preferences, to demographics, natural resources, monetary policy or other instruments. On the other hand, volatility is often attributed to changes in opinions based on news, or psychology connected to the confidence or speculative enthusiasm. 

Uncertainties about the outcome of investment disputes is the key towards understanding volatility of shares of firms involved in disputes that we aim to study.\footnote{It could be argued that poor judgement on the part of investors in the market could be another explanation of volatility (e.g. \cite{shiller2015irrational,simon1991bounded,kahneman2003maps}), \cite{fisman2019event} have recently proposed a method of capturing the imperfections in in the investment behavior. In other words, one could theoretically distinguish between uncertainty about investment dispute and uncertainty pertaining the mind and decisions making of investors. We clearly are working with the former rather than the latter. } We view this uncertainty as an unusual driver of volatility in excess of ``normal'' market uncertainty and volatility associated with it as experienced by investors. Hence, as a proxy for the usual price fluctuations, we use broad S\&P 500 index and we assume that it will help us control all the economic and psychological effects. In addition, it is a widely accepted stylized fact that volatility is dynamically changing over time \cite{engle1982autoregressive,bollerslev1986generalized} hence, by controlling for the time dynamics, we aim to test the excess volatility around the outcome date.

 Following \cite{bialkowski2008stock}, we employ the GARCH model, which allows us to analyze time varying volatility during a period preceding the announcement window. We shall examine the pattern of volatility over a long period of time in which the critical day is the day when the outcome is announced to the market. We make a distinction between disputes won by investors and those that went in favor of host states. We also allow for the magnitude of compensation to see whether and how it affects the volatility. Clearly, an arbitral award that brings in compensation for damages in excess of the amounts claimed by the investor is likely to lead to ``rewards'' by the market and vice versa. In addition, the data matching disputes with share prices are relatively limited. Small samples create statistical problems, and we are pleased that we were able to work with a relatively large sample. 

 The paper is divided into seven sections. The following section 2 provides a brief review of the theoretical literature with the view of identifying specific origins of uncertainties related to arbitrations. Section 3 covers our detailed description of data, its sources and the way the data has been used in our work. Section 4 deals with methodological issues -- especially a detailed description of the GARCH model and the way it is used by us in the paper. The results of estimations are reported section 5. In section 6, we report the results of a robustness test and conclude with final section 7.  

\section{The Theory in the Literature}

Do disputes matter? Do they matter especially for investors? Investment disputes and their effects have been discussed at length in the legal literature and typically only in the context of litigations addressed by courts in the United States. International disputes originating in international trade and investment have so far received far less attention in the literature, and among those disputes, trade disputes have dominated the debate ( e.g. \cite{mavroidis2005wto}). 

For investment disputes to matter for investors, the disputes must affect -- positively or negatively -- the value of the firm as a result of the dispute. From the literature dealing with litigations in the United States we know that litigations/disputes matter for investors because they may indeed affect the value of the firm \citep{arena2017survey}. It is only through the understanding of the way disputes are resolved that the market can assess the risks emanating from the dispute for the investor. 

Ultimately, the question about disputes is -- what is the value of awards?  Until the decision is taken, the value is uncertain. The value will, first of all, depend on whether the  investors will be at all able to receive compensation for his/her loss of current and future income. Note that damages in the ISDS system cover not only losses of income in the past but also losses coming from expected income in the future.  Compensations in investment disputes are provided in the form of cash payment rather than retaliation as it is the case in trade disputes. This means that the award has a clear monetary value -- in contrast to awards in the form of retaliations where the  value is far more complex and ambiguous. On the other hand, the fact that damages in investment disputes  are assessed as past losses as well as the value of future harms. This makes the evaluation of the damages ambiguous for different reasons. The arbiters must have a solid command not only over highly technical details of markets and the balance sheets of the investors but they will also have to agree on the costs of money and other matters of discounting.\footnote{For other differences see \cite{ossa2020disputes}.}

The list of complications in assessing the value of awards does not end with the description of challenges for economists and accountants. The legal literature points to another issue important for evaluating cash payments: quality of courts and their ability to assess damages. Clearly, courts in different jurisdictions do work differently and differences in their judgements cannot be excluded. 

Perhaps  the most complex issue related to the evaluation of awards concerns the likely effect of the award on welfare of the plaintiff and the defendant state. There are two, basically opposing views about the effect. One view, originating in the legal literature, is that litigation leads to welfare losses to both plaintiff and the defendant.\footnote{As noted by \cite{brada2020two}, the general premise of many studies is that there are joint wealth losses for the plaintiff and the defendant corporations but that the biggest change in value is experienced by the firm being sued \citep{bhagat1994costs,bizjak1995effect,bhagat1998shareholder}. By winning the dispute, investor will reduce his losses. Losing the disputes, his losses will augment.} The alternative view is that international arbitrations may, in fact, help increase welfare. It will do so through the instrument of IIA which are seen as a tool of increasing efficiency and hence welfare.\footnote{See. \cite{ossa2020disputes,kohler2019,horn2019economics}.}  

The critics of international arbitration explain the welfare benefits for investors yet from another perspective. For example, for \cite{brada2020one}, international arbitrations are less risky than litigations in national courts (presumably US) and they are biased....\footnote{To quote \cite{brada2020one}, ``The pool of arbitrators is made up mostly of individuals from developed countries, which are likely to be the homes of many MNCs, while respondents are frequently developing countries. If there is some arbitrator bias in favor of the legal systems of developed countries or of the rights of investors from developed countries, there is the possibility that plaintiffs are likely to obtain more favorable awards due to such bias on the part of the arbitra-tors. Of course, the fact that arbitrators may reduce awards below the amounts claimed by plaintiffs does not disprove that they may favor foreign investors in their decisions because plain-tiff investors have strong incentives to overstate their claims, both to get large judgments and to encourage respondent states to settle claims rather than to risk large losses in arbitration. Thus, even if arbitrators do reject some of the damages claimed by the plaintiffs, their awards may still be more favorable to plaintiffs than the law and the facts of the case warrant.''}   This argument  echoes similar views that have been expressed in the literature on trade disputes (e.g. \cite{horn2005use}). The differences suggest that arbitration under a BIT carries less risk and possibly greater rewards for the plaintiff than does inter-firm litigation in national courts. These views are more or less shared by other critics such as \cite{sachs2019investment,van2011use,rodrik2017straight}  and others.\footnote{It is not clear why costs and benefits of national litigation are comparable with costs and benefits of international arbitrations when we discuss the effects of the latter on investors. Investors are not typically facing a choice between litigation in their  national courts or in national courts abroad. International arbitration is their only option in those cases.} In contrast, investment disputes also generate uncertainties about the International System of Dispute Settlement (ISDS) . To begin with, ISDS of investment disputes is different from that used in the resolution of trade disputes \citep{ossa2020disputes}, they are subject to different provisions of IIA and those provisions have been evolving over time \citep{pohl2018societal}. This makes it more complicated for market participants to evaluate and anticipate  the outcomes of arbitrations. International arbitration can be subject to different interpretations of agreed rules, and this, again, increases the uncertainty and unpredictability of arbitrations. The jurisprudence of international investment law is evolving and may sometimes lead to different interpretations of the same rule \citep{howselevin2020}.

The differences in the views about international arbitration have serious implications for economists and their attempts to assess empirically the effect of disputes on investors. The different perceptions about expected welfare gains complicate the choice of the dependent variable to be measured -- especially the choice between the absolute level of real rate of return and volatility of stock. Given the theoretical differences in the expected changes of welfare, the expected level of the real rate of return is ambiguous. Does the rate have to be necessarily above a benchmark or would some other values be still be acceptable? This ambiguity is  one of the reasons for the choice of  volatility as our dependent variable in this study. Other benefits of international arbitration to investors can also be noted. They range from the value of government commitments to protect future profits of investors to benefits from various signaling functions of government commitments or the benefits from a better control of  moral hazard.

However, the risks to investors are also considerable. The ``classical'' problem faced by investors is  the \textit{expropriation risk}, leading to costs of what is known as  police carve--outs \citep{aisbett2010compensation}: Police carve/outs represent a range of regulations giving the host country government certain degree of flexibility and allowing it to address a policy issue. The regulation may lead to moral hazard and inefficiencies but a more serious problem is arguably the fact that they represent a potential ``hold up'' problem. Once the investment is sunk, the expropriation of assets is a ``gain'' for the host country \textit{in addition} to  the value the country generates for itself by addressing the public policy issue.\footnote{Much of the recent literature on optimal investment agreements looks at ways of designing an optimal compensation mechanism in cases of indirect expropriation under IIA. For a review of the literature, see \cite{aisbett2010compensation}.}

Another reason for high costs of investors are excessive \textit{regulatory} takings.\footnote{The literature on regulatory takings originates in the paper by \cite{blume1984taking} and it would be outside scope of this paper to review it.} Host country government's regulations may be ``weak'' or ``strong''. In the latter case, we say that regulatory takings are excessive because they disregard the (welfare) interests of investors. As noted by \cite{aisbett2010police},the lack of regard for the investor's welfare may lead to inefficient over-regulation by the host even in the absence of a ``hold-up'' motivation. The same argument is made by \cite{horn2019economics} in their paper on optimal investment agreements. 

The third reason why the costs of investors may be understated is the fact that the costs of host country's regulations are seen by host country governments as being ``external''. In other words, host country governments do not typically internalize the costs that local regulations impose on FDI. Those costs have to be born by the investor and even though the investor may attempt to pass the costs onto consumers in the country they may not be able to do so due to competition.\footnote{As pointed out by \cite{aisbett2010police}, cost-internalization is the leading justification among legal scholars for the U.S. Fifth Amendment, which states that private property shall not be “taken for public purpose without just compensation”. } 

The fourth risk faced by investors is the enforceability of the arbiters' decision even when the decision is in their favor. In theory, the decision can be enforced only if the host country has the resources to pay for the damages and when its acceptance of the decision is driven both by its respect for the court and by its concerns about its reputational costs. The latter could be high and affect the host country's attractiveness to future business. Even though the costs of sovereign defaults have lately increased \cite{schumacher2018sovereign}, and hence their willingness to comply with the decision of the court, the risk of non-payment for damages remains high. 

Finally,  there are other costs that need to be mentioned such as  currency risk that can be significant in countries with unstable currencies. National competitors do not normally take such risks. In addition,  considering differences in linguistic, cultural and historical backgrounds, it is arguable whether the competitive conditions in the market are on a  level playing field even if national treatment and ( and MFN) are formally a part of the conditions of establishment and operations. The reason is that NT and MFN provisions primarily  focus on taxes and other financial instruments and completely disregard the role of social, cultural  and historical factors.  Moreover, the legal language of IIA is often vague. E.g. The NT provision, which calls for the same conditions ``in like circumstances'' is clearly  one example of many that have created many controversies in the WTO DSB mechanism and in international arbitrations.\footnote{\cite{aisbett2010police} makes a similar point when they argue that ``.. the treaties' investor to state provision and the use of international tribunals create an asymmetry between domestic and foreign investors, causing their circumstances to be inherently ``unlike''. Thus, there is a tension between the expropriation and the non-discrimination clauses.''}

In conclusion, the adjudication of claims made by investors and the evaluation of awards are clearly a daunting task for arbiters of investment disputes. The task is complicated from moment when the arbiters have to decide whether and which investor's property right has been violated. They then have to demonstrate that the right has been violated by the host country policy intervention, and they have to establish whether the intervention was within the police carve out provisions. Finally, they must have a notion of the amount of damages that can be reasonably offered to the investor once the claim has been accepted. The tasks involve many unknowns. As the outcome of arbitration is uncertain so will have to be the expected impact on the firm's value, and until the decision is taken, the uncertainty must be translated into uncertainty about the valuation of the firm by the market.

\section{Data About International Investment Disputes} \label{data}

The research on investment treaty arbitration lacks a comprehensive data set of all disputes filed or resolved. Transparency, the quality and public availability of investment disputes  is improving but the data must still be read with caution.  There are various sources of data on investment disputes - some are related to data on arbitrations produced by Oxford University Law Center and published by Oxford University Press\footnote{OUP Press Investment Claims: \url{http://oxia.ouplaw.com}}or by Kluwer Arbitration,\footnote{Kluwer Arbitration : \url{http:// kluwerarbitration.com}} other sources originate in data on IIA which cover, inter alia, information about dispute settlement mechanism. The most frequently used  sources of  data on investment arbitration  are documents released by the United Nations Conference on Trade and Development\footnote{\url{https://investmentpolicy.unctad.org/investment-dispute-settlement}} (UNCTAD), and International Centre for Settlement of Investment Disputes\footnote{\url{https://icsid.worldbank.org/en/Pages/cases/AdvancedSearch.aspx}} (ICSID). UNCTAD reports only ``known'' cases and will, therefore, exclude cases that are not publicly known. In contrast, the ICSID statistics are also under-inclusive   because they only rely on data from  cases covered under the ICSID jurisdiction.\footnote{For more discussion of differences between both sources, see \cite{shirlow2015}.} According to one estimate, the ICSID cover about 70 percent of all disputes. Thus, the UNCTAD data are more ``inclusive'' than the ICSID data but are also likely to be incomplete. In addition, neither of the two sources covers each reported case with full information about the arbitral process, and we had to combine the two sources. Despite the weaknesses, the documents are the best efforts representation of trends in investment arbitration and should not significantly skew our analysis. There is also a recent effort to provide a comprehensive list of all known cases in the form of publicly available dataset from PITAD Investment Law and Arbitration Database.\footnote{\url{https://pitad.org/index\#static/about_pitad}. See also \url{https://www.acerislaw.com/choosing-icsid-or-uncitral-arbitration-for-investor-state-disputes/}} This source aims at providing an exhaustive data set of all cases but, on the other hand, it lacks some of the information regarding the individual cases which are provided in other sources such as UNCTAD or ICSID. 

We work with the versions of the documents from  January 2019. We scraped the data from the respective web pages and had to merge the information when constructing our database in order to obtain the information necessary for our estimations. On the one hand, the ICSID database contains more detailed information regarding some of the variables. For example, it contains the exact date of the registration of the dispute, which is needed as a control date in our experiments. The information is not available in the UNCTAD source. On the other hand, ICSID does not explicitly contain information regarding the final decision of the court and the amount claimed/awarded, another important piece of information in our analysis. As a check of our merged data, we verified that the key information contained in both databases such the outcome date match in both sources. Table \ref{tab:cases} provides a list of all cases that are included in the analysis. In total, we work with a sample of 46 cases that resulted in a decision  in favor of either party or were settled or discontinued. Out of these 46 cases, 25 cases were decided in favor of the investor, 10 in favor of the state, and 11 were either settled or discontinued.  

We should note that while we were able to extract information about a large number of companies from the combination of the two documents,  only a small part of the data  could finally be used in the analysis. The reason is that we needed to match the data from both sources with stock market data of the companies. Unfortunatelly, the number of companies involved in disputes whose shares were also traded in stock markets was relatively limited. The limitations on the availability of the liquid price information of the claimant (an individual or company) constrained the size of our sample. Most of the investment agreements were BITs between the investor's and the host countries, but the claims included in our sample were also brought under NAFTA, CIS Investor Rights Convention and the Energy Charter Treaty provisions. Our sample covers the period in which the year of initiation of dispute resolution ranges between 1998 and 2014, and the year of decision (outcome of the arbitration) is  between 2000 and 2017. The investments disputes cover  all economic sectors (primary 10, secondary 8, and tertiary). Table \ref{tab:cases} also contains information about the investor's country (highest number from the United States - 12), the host country (highest number from Argentina - 10), and the amount claimed by the investor and amount rewarded by the tribunal.\footnote{The size of the sample was constrained by a number of other factors. In some arbitral proceedings, the arbitrators declined jurisdiction over the dispute. Many cases could not be used in our sample because the plaintiffs (individuals or firms) were not listed on any stock exchange.  Moreover, MNC often use affiliates as a vehicle for their FDI,  and those affiliates are also not listed in stock exchanges.} With the exception of China, all of the investor's countries are developed countries. As for the host countries, seven cases come from transition countries and thirty three from emerging markets.

Information regarding the stock prices was obtained from Yahoo Finance.\footnote{\url{https://finance.yahoo.com}} In order to isolate the effect of investment dispute resolution from general market movements, we removed the effect of the aggregate market from the stock price of the investor's firm. We use the S\&P500 index as a proxy for general market changes because  most of the companies in our sample operate internationally  and the S\&P500 market is a good proxy for the stock market globally. By using one single benchmark, we also limit the effect of differences in stock price volatility in different securities markets. Stock prices employed are either direct prices of the company  that made the investment, or prices of the parent's company.

\begin{sidewaystable}
\centering
\tiny
\caption{List of cases included in the analysis.}
\label{tab:cases}
\begin{tabular}{cclllllrr}
  \toprule
\tiny
\thead{Registration \\ year} & \thead{Outcome \\ date} & \thead{Plaintiff} & \thead{Applicable \\ IIA} & \thead{Decided \\ in favour of} & \thead{Host \\ country} & \thead{Nationality \\ of investor} & \thead{Amount \\ claimed \\ (mil US\$)} & \thead{Amount \\ awarded \\ (mil US\$)} \\ 
  \midrule
2005 & 2008-11-12 & Astaldi, S.p.A. & Algeria - Italy BIT (1991) & State & Algeria & Italy &  & 0.00 \\ 
  2007 & 2016-12-21 & HOCHTIEF Aktiengesellschaft & Argentina - Germany BIT (1991) & Investor & Argentina & Germany & 157.20 & 13.40 \\ 
  2007 & 2011-06-21 & Impregilo S.p.A. & Argentina - Italy BIT (1990) & Investor & Argentina & Italy & 119.00 & 21.29 \\ 
  2003 & 2015-04-09 & Suez, S.A. & Argentina - France BIT (1991) & Investor & Argentina & France & 834.00 & 383.60 \\ 
  2002 & 2007-09-28 & Sempra Energy International & Argentina - United States of America BIT (1991) & Investor & Argentina & USA & 209.00 & 128.00 \\ 
  2002 & 2007-02-06 & Siemens A.G. & Argentina - Germany BIT (1991) & Investor & Argentina & Germany & 462.50 & 237.80 \\ 
  2001 & 2005-05-12 & CMS Gas Transmission Company & Argentina - United States of America BIT (1991) & Investor & Argentina & USA & 261.10 & 133.20 \\ 
  2005 & 2012-08-22 & Daimler Financial Services AG & Argentina - Germany BIT (1991) & State & Argentina & Germany & 243.00 & 0.00 \\ 
  2012 & 2014-05-19 & Repsol, S.A. & Argentina - Spain BIT (1991) & Settled & Argentina & Spain & 10500.00 & 5000.00 \\ 
  2008 & 2010-10-27 & Impregilo S.p.A. & Argentina - Italy BIT (1990) & Settled & Argentina & Italy & 250.00 &  \\ 
  2004 & 2006-03-29 & France Telecom S.A. & Argentina - France BIT (1991) & Settled & Argentina & France & 300.00 &  \\ 
  2005 & 2009-06-30 & Saipem S.p.A. & Bangladesh - Italy BIT (1990) & Investor & Bangladesh & Italy & 12.50 & 6.30 \\ 
  2012 & 2015-04-30 & Ping An Insurance (Group) Company of China, Limited & BLEU - China BIT (1984, 2005) & State & Belgium & China & 960.60 & 0.00 \\ 
  2007 & 2015-02-20 & Murphy Oil Corporation & NAFTA & Investor & Canada & USA & 59.10 & 13.90 \\ 
  2006 & 2012-10-05 & Occidental Petroleum Corporation & Ecuador - United States of America BIT (1993) & Investor & Ecuador & USA & 1000.00 & 1769.00 \\ 
  2006 & 2008-12-01 & Chevron Corporation & Ecuador - United States of America BIT (1993) & Investor & Ecuador & USA & 649.00 & 77.70 \\ 
  2003 & 2006-02-03 & EnCana Corporation  & Canada - Ecuador BIT (1996) & State & Ecuador & Canada & 80.00 & 0.00 \\ 
  2013 & 2016-12-13 & Edenred, S.A. & France - Hungary BIT (1986) & Investor & Hungary & France &  & 24.30 \\ 
  2009 & 2012-06-11 & Electricite de France (EDF) International S.A. & The Energy Charter Treaty (1994) & Investor & Hungary & France & 100.00 & 132.60 \\ 
  2004 & 2006-09-13 & Telenor Mobile Communications AS & Hungary - Norway BIT (1991) & State & Hungary & Norway & 152.00 & 0.00 \\ 
  2007 & 2010-09-23 & AES Summit Generation Limited & The Energy Charter Treaty (1994) & State & Hungary & United Kingdom & 230.00 & 0.00 \\ 
  2001 & 2003-10-07 & AIG Capital Partners, Inc. & Kazakhstan - United States of America BIT (1992) & Investor & Kazakhstan & USA & 13.50 & 6.00 \\ 
  2013 & 2014-06-30 & Stans Energy Corp. & CIS Investor Rights Convention (1997) & Investor & Kyrgyzstan & Canada & 117.80 & 117.80 \\ 
  2002 & 2005-02-22 & France Telecom & France - Lebanon BIT (1996) & Investor & Lebanon & France & 952.00 & 266.00 \\ 
  2009 & 2011-09-02 & EVN AG & The Energy Charter Treaty (1994); Austria - Macedonia BIT (2001) & Settled & Macedonia & Austria & 229.10 &  \\ 
  2009 & 2013-04-18 & Abengoa, S.A. & Mexico - Spain BIT (2006) & Investor & Mexico & Spain & 70.00 & 40.30 \\ 
  2004 & 2007-11-21 & Archer Daniels Midland & NAFTA & Investor & Mexico & USA & 100.00 & 33.50 \\ 
  2004 & 2009-08-18 & Corn Products International, Inc. & NAFTA & Investor & Mexico & USA & 325.00 & 58.00 \\ 
  1998 & 2000-06-02 & Waste Management, Inc. & NAFTA & State & Mexico & USA & 36.00 & 0.00 \\ 
  2000 & 2004-04-30 & Waste Management, Inc. & NAFTA & State & Mexico & USA & 36.60 & 0.00 \\ 
  2004 & 2007-11-21 & Tate \& Lyle Ingredients Americas, Inc. & NAFTA & Investor & Mexico & USA & 100.00 & 33.50 \\ 
  2003 & 2005-09-26 & Impregilo S.p.A. & Italy - Pakistan BIT (1997) & Settled & Pakistan & Italy & 450.00 &  \\ 
  2014 & 2017-11-30 & Bear Creek Mining Corporation & Canada-Peru FTA & Investor & Peru & Canada & 522.20 & 18.20 \\ 
  2012 & 2014-03-25 & Elecnor, S.A. & Peru - Spain BIT (1994) & Settled & Peru & Spain &  &  \\ 
  2003 & 2007-08-16 & Fraport AG Frankfurt Airport Services Worldwide & Germany - Philippines BIT (1997) & State & Philippines & Germany & 425.00 & 0.00 \\ 
  2002 & 2008-04-11 & SGS Société Générale de Surveillance, S.A. & Philippines - Switzerland BIT (1997) & Settled & Philippines & Switzerland & 140.00 &  \\ 
  2009 & 2012-10-31 & Deutsche Bank A.G. & Germany - Sri Lanka BIT (2000) & Investor & Sri Lanka & Germany & 60.00 & 60.00 \\ 
  2002 & 2007-07-19 & Canfor Corporation & NAFTA & Settled & United States of America & Canada & 250.00 &  \\ 
  2010 & 2016-07-08 & Philip Morris Products, S.A. & Switzerland - Uruguay BIT (1988) & State & Uruguay & Switzerland & 22.30 & 0.00 \\ 
  2012 & 2017-11-03 & Saint-Gobain Performance Plastics Europe & France - Venezuela BIT (2001) & Investor & Venezuela & France & 115.10 & 34.40 \\ 
  2011 & 2016-01-29 & Tenaris, S.A. & BLEU - Venezuela BIT (1998) & Investor & Venezuela & Luxembourg & 299.30 & 87.30 \\ 
  2010 & 2014-11-18 & Flughafen Zürich, A.G. & Switzerland - Venezuela BIT (1993) & Investor & Venezuela & Switzerland & 82.20 & 19.40 \\ 
  2007 & 2014-10-09 & Exxon Mobil Corporation & Netherlands - Venezuela BIT (1991) & Investor & Venezuela & USA & 14679.00 & 1600.00 \\ 
  2012 & 2012-11-29 & Ternium, S.A. & BLEU - Venezuela BIT (1998) & Discontinued & Venezuela & Luxembourg & 130.00 &  \\ 
  2011 & 2017-02-08 & The Williams Companies, International Holdings B.V. & Netherlands - Venezuela BIT (1991) & Settled & Venezuela & United Kingdom &  &  \\ 
  2008 & 2012-02-15 & CEMEX Caracas Investments B.V. & Netherlands - Venezuela (1991) & Settled & Venezuela & Netherlands & 1200.00 & 600.00 \\
   \bottomrule   
\end{tabular}
\end{sidewaystable}

\newpage
\section{Measurement of Abnormal Volatility Around the Outcome Date}

Volatility, as a proxy for the uncertainty in the markets, has been one of the most important phenomena studied in financial economics.  In his seminal work on stock market volatility, \cite{shiller1981stock} argues that the observed volatility is not consistent with the predictions of the present value models and that the high intertemporal variation cannot be rationalized on that basis. The failure of standard valuation models to explain the high magnitudes of stock market fluctuations opened serious challenge to financial  economists, and literature started the search for the volatility drivers different from conventional dividends and earnings. One strand of the literature examines whether the variability of returns can be linked to macroeconomic variables, financial leverage, or trading volume (e.g. \cite{schwert1989does,binder2001stock,engle2013stock}).

A different route is explored by event studies connecting the excess volatility with some event.\footnote{The approach also involves less stringent assumptions than those required in standard event studies. The latter typically look at the effect of investment disputes on the \textit{absolute} level of real returns. To do so, they need to consider the relationship between the absolute level of award in arbitration and the change in the firm's value in order to assess the real damages to investor. This strict assumption is not necessary in studies of volatility. Yet another approach has recently been suggested -- judgement of investors in the market. As \cite{fisman2019event} suggest, investors' poor judgement may be the reason for volatility. We assume in this paper that poor judgement does not characterize \textit{all} market participants and that individual mistakes do not dominate the market.} From early efforts summarized by \cite{yadav1992event}, researchers subsequently studied the reaction of volatility to political events such as national elections \citep{bialkowski2008stock}, mergers and acquisitions \citep{balaban2006volatility} or terrorist attacks \citep{essaddam2015event}. Abnormal volatility, to the best of our knowledge, has not yet been connected to the investment disputes.

To account for strong time variation in volatility, we need to step forward from traditional event study techniques. Estimating the model that will capture the time varying volatility during the window period immediately preceding the outcome window and including  the market volatility as control  variable for volatility induced by market, we will first estimate an expected, or an usual volatility under the null hypothesis of no reaction of the market to the event.  Then to see if the resolution of the dispute induced changes in the volatility, we will compare this expectation to the measured volatility in the time window around the outcome date. The key aspect of the methodology used is to capture the usual time varying volatility, which allows us to measure abnormal variation in response to the event. Moreover, following \cite{bialkowski2008stock}, we construct a bootstrap procedure to our tests in order to mitigate small sample bias.

Let us discuss the procedure in detail. The key date is the outcome date $t_{\text{O}}$.  The announcement date of the investor's claim is not taken as a ``surprise date'' as market will have the information already processed from media. The surprise cannot be measured and timed, but the  outcome day can be timed. The data are measured over the time period $t\in[1,\ldots,T]$  so that  $t=1$ is the starting date and $t=T$ is end of that period, so that outcome date $t_{\text{O}}$ is inside of the interval defined as $1<t_{\text{O}}<T$. This allows us to capture the specifics of each stock.\footnote{Note that the dates of claims and the dates of decisions differ from case to case. Note also that the announcement is given as an announcement window, which we define further below.} Since uncertainty about the outcome will probably result in an abrupt volatility changes even before the announcement day, we focus on the time window  around the outcome date that includes certain period of time before  as well as after the outcome. Formally, the window of interest is defined as $[t_{\text{O}}-N_{\text{B}},t_{\text{O}}+N_{\text{A}}]$ with $N_{\text{B}}$ and $N_{\text{A}}$ being a fixed number of days before and after the outcome day $t_{\text{O}}$ respectively, and we will refer to it as an announcement window.\footnote{In real world, the announcement window could be ``contaminated'' by ``other events''. We proxy the contamination by ``other events'' by looking at the excess volatility controlled by returns (and volatility as well) of  S\&P500 index and assume that S\&P500 controls for this effect as we discuss the best practice option in the text.}

We start by estimating the case-specific component of variance for all companies in the sample $i=1,\ldots,K$ with generalized auto-regressive conditional heteroscedasticity (GARCH) model over time period $t\in[t_{\text{O}}-N_{\text{B}}-500,\ldots,t_{\text{O}}-N_{\text{B}}]$ immediately before the event window as 
\begin{eqnarray}
r_{i,t} &=& \alpha + \beta r_{t}^{\text{S\&P 500}}+\epsilon_{i,t} \label{eq:garch1} \\
\epsilon_{i,t} &\sim & \mathcal{N}(0,\sigma_{i,t}^2) \label{eq:garch2}\\ 
\sigma_{i,t}^2 &=& \psi_0 + \psi_1 \sigma_{i,t-1}^2 + \psi_2 \epsilon_{i,t-1}^2 \label{eq:garch3}
\end{eqnarray}
where $r_{i,t}$ is logarithmic return of the $i$th company and $r_{t}^{\text{S\&P 500}}$ is stock market return represented by the S\&P 500 market index.

The model described by equations \ref{eq:garch1} - \ref{eq:garch3} is a GARCH(1,1) model and is one of the most popular and useful frameworks for capturing the time varying volatility. \cite{engle1982autoregressive,bollerslev1986generalized} noted that returns described by equation \ref{eq:garch1} display strong time variation in its second moment, volatility, and proposed to describe it by latent dynamics in \ref{eq:garch3} assuming the residuals \ref{eq:garch2} being normally distributed.\footnote{Note that GARCH(1,1) is powerful benchmark that captures the time variation in volatility well. Large literature trying to extend the model in various ways does not find significant improvements of out-of-sample forecasts. An interesting survey of \cite{hansen2005forecast} summarizes and compares 330 extensions and find no evidence that GARCH(1,1) is outperformed by more sophisticated models. Hence we chose the model as a best parsimonious choice.} We chose to estimate the model for the period of 500 days immediately before the announcement window. This should allow the period  to be reasonably long in order for the market to absorb the information and to address the possibility of the  small sample bias \citep{hwang2006small}.

Being interested in measurement of abnormal volatility, we consider variation in $\epsilon_{i,t}$ around the outcome date  (i.e. outcome window) in relation to the usual volatility estimated in the period before the outcome window. In other words, model described by equations \ref{eq:garch1} -- \ref{eq:garch3} will indicate the level of expected usual volatility under the null hypothesis of no reaction to the announcement of the resolution. Comparison of the modelled volatility to the volatility measured by the actual data will then indicate the abnormal volatility. Abnormal volatility is further controlled for stock market volatility. Note that this benchmark volatility captures time dynamics in the volatility as well as volatility induced by S\&P500 that is by estimating $\beta$ coefficient, and including  market returns into the model directly. Formally, we define it as a $k$-step-ahead  forecast of the variance conditional on the information available on the last day of the period before the announcement window
\begin{equation}
\mathbb{E}_t\Big[\sigma_{i,t+k}^2|\mathcal{I}_t\Big] = \widehat{\gamma}_0\sum_{j=0}^{k-1} (\widehat{\gamma}_1+\widehat{\gamma}_2)^j + (\widehat{\gamma}_1+\widehat{\gamma}_2)^{k-1} \widehat{\gamma}_1 \sigma_{i,t}^2 + (\widehat{\gamma}_1+\widehat{\gamma}_2)^{k-1} \widehat{\gamma}_1 \epsilon_{i,t}
\end{equation}
Importantly, the distribution of the residuals during the announcement window can be described as $\epsilon_{i,t} \sim \mathcal{N}(AR_t,\mathcal{M}_t \mathbb{E}_t\big[\sigma_{i,t+k}^2|\mathcal{I}_t\big])$ where the residuals will on average hold abnormal return $AR_t$, and $\mathcal{M}_t$ is the multiplicative effect of the announcement-induced volatility. Under the null hypothesis that investors are not surprised by outcome of the dispute,\footnote{Note that we are making a distinction between two situations to be considered by the market -- event or no event. We do not differentiate among different outcomes of the event. The assumption seems reasonable since it is impossible for the market to predict the precise outcome of the arbitration.} the parameter $\mathcal{M}_t = 1$. Given our interest in capturing the effect of  investment disputes on volatility, announcement-induced volatility $\mathcal{M}_t$ is of primary  interest and needs to be estimated.  By noting that residuals demeaned  using cross-sectional average are normally distributed with zero mean and announcement-independent variance, we can estimate the $\mathcal{M}_t$ as the cross-sectional variance of demeaned residuals by announcement-independent standard deviation \citep{boehmer1991event}
\begin{equation}
\widehat{\mathcal{M}}_t = 1/(K-1)\sum_{i=1}^K\Bigg(\frac{\Big(K\widehat{\epsilon}_{i,t} - \sum_{j=1}^K \widehat{\epsilon}_{j,t} \Big)^2 }{K(K-2)\mathbb{E}_t\Big[\sigma_{i,t}^2|\mathcal{I}_t\Big] + \sum_{j=1}^K \mathbb{E}_t\Big[\sigma_{j,t}^2|\mathcal{I}_t\Big]} \Bigg)
\end{equation}
with $\widehat{\epsilon}_{i,t}=r_{i,t} - \widehat{\alpha} - \widehat{\beta} r_{t}^{\text{S\&P 500}}$ and $t\in [t_{\text{O}}-N_{\text{B}},t_{\text{O}}+N_{\text{A}}]$.

Since residuals follow standard normal distribution under the null hypothesis when  $\mathcal{M}_t = 1$, abnormal change in volatility is $\mathcal{M}_t - 1$, and hence the \textbf{cumulative abnormal volatility} $\mathcal{CAV}\Big(N_B,N_A\Big)$ can be computed as
\begin{equation}
\mathcal{CAV}\Big(N_B,N_A\Big) = \Bigg(\sum_{t=t_{\text{O}}-N_{\text{B}}}^{t_{\text{O}}+N_{\text{A}}}\widehat{\mathcal{M}}_t\Bigg) - (N_A-N_B+1).
\end{equation}

The null hypothesis of no impact of investment dispute on volatility
\begin{equation}
\mathcal{H}_0: \mathcal{CAV}\Big(N_B,N_A\Big) = 0
\end{equation}
is equivalent to 
\begin{equation}
\mathcal{H}_0: \sum_{t=t_{\text{O}}-N_{\text{B}}}^{t_{\text{O}}+N_{\text{A}}} \mathcal{M}_t (K-1) = (N_A-N_B+1)(K-1),
\end{equation}
and can be tested noting that under the null, $\mathcal{M}_t$ is a variance of $K$ independent $\mathcal{N}(0,1)$ random variables implying the test statistic for the hypothesis is $\chi^2$ distributed as
\begin{equation}
\sum_{t=t_{\text{O}}-N_{\text{B}}}^{t_{\text{O}}+N_{\text{A}}}(K-1)\widehat{\mathcal{M}}_t \sim \chi^2_{(K-1)(N_B-N_A+1)}.
\end{equation}
Moreover, inference based on the asymptotic values may be biased in case of non-normally distributed residuals, cross-sectional dependence, or autocorrelation structures that are likely to be observed in the data. As suggested by \cite{bialkowski2008stock}, we additionally construct a bootstrap test based on \cite{efron1992bootstrap}. The cumulative abnormal volatility during the announcement window is compared with the empirical distribution of $\mathcal{CAV}\Big(N_B,N_A\Big)$ simulated under the null hypothesis. The procedure for generating the empirical distribution can be described as follows:
\begin{enumerate}
	\item Draw the case and date to match the number of outcomes in the original sample randomly with replacement from the whole dataset.
	\item Compute $\mathcal{CAV}\Big(N_B,N_A\Big)$ from the randomly generated sample from step (1).
	\item Repeat step (1) and (2) 5000 times, sort the estimated $\mathcal{CAV}\Big(N_B,N_A\Big)$ in order to obtain empirical distribution.
\end{enumerate}
The bootstrapped $p$-value is then computed as the number of bootstrapped $\mathcal{CAV}\Big(N_B,N_A\Big)$ that exceed the $\mathcal{CAV}\Big(N_B,N_A\Big)$ calculated on the original election date, divided by the number of replications. Note this is one-sided test to test the null hypothesis against the alternative $\mathcal{H}_A: \mathcal{CAV}\Big(N_B,N_A\Big) > 0$, but in case the number of bootstrapped statistics below the value of statistics estimated on data is used, alternative hypothesis $\mathcal{H}_A: \mathcal{CAV}\Big(N_B,N_A\Big) < 0$ can be tested. In other words, we can confirm if the potential volatility connected to the dispute is increasing or decreasing.

\section{Results: Return Volatility around the Outcome of the Investment Dispute}

We chose the outcome date to be the main event for the choice of the announcement window. We divide the sample into cases in which  the decision was  made in favor of the state, cases with the decision in favor of investor, as well as cases that have been discontinued or settled. These three groups are distinct in their potential impact of the decision on the  plaintiff company, and hence we expect a very different reaction of investors. While our expectation is that the decision in favor of state will result in increased volatility, capturing increased uncertainty about the prices since the decision may lead to  large costs for the company. On the other hand, cases with decisions in favor of investor as well as those that were settled are likely to show the opposite pattern of volatility. 

\begin{figure}[ht!]
  \begin{center}
\caption{\textbf{Volatility reaction to the Dispute Outcome}} \label{fig:CAV}
\includegraphics[width=0.8 \textwidth]{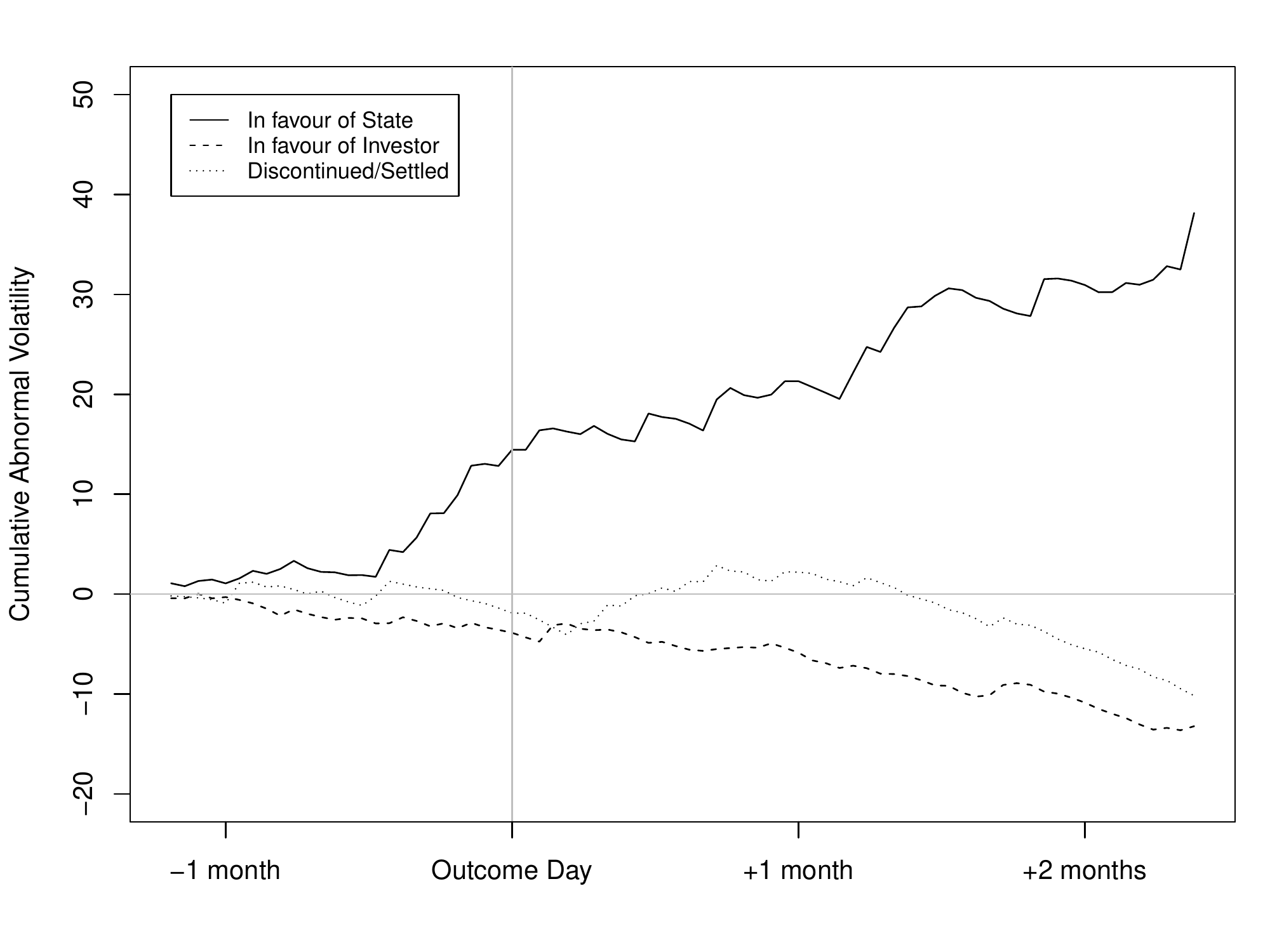}
\caption*{\scriptsize \textit{Notes}: The figure shows cumulative abnormal volatility $\widehat{\mathcal{CAV}}\Big(N_B,N_A\Big)$ around the outcome day. Reaction of companies with positive outcome are captured by dashed line, reaction of companies with negative outcome (in favor of state) in black, the ones settled out of court or undecided are in dotted line. The window is one month before and two months after the announcement day.}
\end{center}
\end{figure}

Figure \ref{fig:CAV} shows the estimates of the cumulative abnormal volatility $\widehat{\mathcal{CAV}}\Big(N_B,N_A\Big)$ for the three samples. We choose to include one month before the outcome date as well as two months after the outcome day in the announcement window to see the  increased uncertainty before the announcement date.

As can be seen from Figure \ref{fig:CAV}, we document a surge in the volatility of cases decided in favor of state. The increase in the volatility  started even before the outcome date ( about two weeks prior to the date of the outcome), presumably in the anticipation of the results or news leaks. After the outcome day, the volatility continues to surge steadily every single day throughout the  two months period and even beyond. A very different pattern is documented for the cases when the decision has been made in favor of investor. Here the volatility decreases gradually. Cases that have been settled or discontinued show no abnormal volatility while volatility decreases slightly at the end of the reaction window.

\begin{table} [ht!]
  \centering
  \caption{Cumulative abnormal volatility around Decision Day}   \label{tab:decision}
  \begin{threeparttable}
  \footnotesize{
  \centering
  \begin{tabular}{lrrrr}
  \toprule
  Announcement window &$\widehat{\mathcal{CAV}}$ & \% vol &$p$-val & $p$-val boot\\
   \midrule
   \multicolumn{5}{l}{\textit{Panel A: Decision in favour of Investor}}\\
	(-2 days,2 days) & -1.010	&  -0.202 & 0.054 & 0.118 \\
 	(-1 week,1 week) & 2.863 &  0.260 & 0.003 & 0.083 \\
	(-2 week,2 weeks) & 1.624 & 0.077 & 0.116 & 0.327 \\
	(-1 month,1 month) & -7.387&  -0.145 & 0.000 & 0.003 \\
	(-1 month,2 months) & -13.190 & -0.174 & 0.000 & 0.000 \\
	\\
  \multicolumn{5}{l}{\textit{Panel B: Decision in favour of State}}\\
	(-2 days ,2 days) & 1.141 &0.228 &0.129 &0.337 \\
	(-1 week,1 week) & 1.686 &0.153 &0.129 &0.402 \\
	(-2 weeks,2 weeks) & 27.653 &0.317 &0.000 &0.000 \\
	(-1 month,1 month) & 19.763 &0.388 &0.000 &0.015 \\
	(-1 month,2 months) & 45.949 &0.605 &0.000 &0.000 \\
	\\
  \multicolumn{5}{l}{\textit{Panel C: Discontinued/Settled}}\\
	(-2 days,2 days) &-1.691 &-0.338 &0.031 &0.055 \\
	(-1 week,1 week) & -3.700 &-0.336 &0.003 &0.007 \\
 	(-2 week,2 weeks) &3.279 &0.156 &0.000 &0.285 \\
 	(-1 month,1 month) & 0.444 &0.009& 0.000 &0.708 \\
	(-1 month,2 months) & -10.859 &-0.143 &0.000 &0.002 \\
\bottomrule
  \end{tabular}
    \caption*{\scriptsize \textit{Notes}: The table contains cumulative abnormal volatility $\widehat{\mathcal{CAV}}\Big(N_B,N_A\Big)$ around the outcome day. Panel A of the table reports results for the cases decided in favor of Investor, Panel B reports the results for the cases decided in favor of state, and Panel C reports cases that were discontinued or settled. The implied percentage change (\% vol) relative to the benchmark, asymptotic $p$-values as well as $p$-values computed on 5000 bootstraps are reported and test the null hypothesis of no reaction $\mathcal{H}_0: \widehat{\mathcal{CAV}}\Big(N_B,N_A\Big)=0$.}}
  \end{threeparttable}
\end{table}

To test the significance of the results, we compute the cumulative abnormal volatility $\widehat{\mathcal{CAV}}\Big(N_B,N_A\Big)$ statistics for different window windows of 2 days, 1 week, 2weeks, 1month and 2 months before as well as after the outcome day. Specifically, Table \ref{tab:decision} displays the cumulative abnormal volatility in the first column, theoretical as well as bootstrapped p-values in the third and fourth columns respectively. In addition, the second column of the Table \ref{tab:decision} holds percentage change of the volatility relative to the benchmark.

The results confirm the significant increase of abnormal volatility of the cases decided in favor of state in longer than 2 weeks period. The percentage volatility increase is growing with the period, and abnormal volatility of cases decided in favor of state cumulated by 60.5\% in the 1 month before -- 2 months after the outcome day window. In contrast, cumulative abnormal volatility in the cases decided in favor of investors decreased over time, resulting in a significant decrease in abnormal volatility. Investors thus reacted to the ``good news'' with smaller than the usual volatility -- by a  decrease of -17.4\%. Discontinued or settled cases show a similar pattern to those decided in favor of investor with decrease in abnormal volatility of -14.5\%. In brief, volatility of the company's shares was increased as uncertainty about the outcome set in.  When the outcome decision was in favor of investor, the uncertainty decreased back to its ``normal'' level, and hence the excess -- abnormal volatility  disappeared.  On the other hand, if the decision was in favor of state, it increased the uncertainty about the impact of the decision on the balance sheet of the company reflecting partly the loss of income of the investor and partly the cost of litigation.

\section{Robustness Tests}

The previous section presented our estimates of abnormal volatility based on the GARCH model, which serves as the benchmark for volatility  without the event. We have isolated the case -- specific component of variance within GARCH. In order to avoid a bias in estimating abnormal volatility, we have estimated the variance conditional to the prior dates of the event. To recall, the variance was estimated by the event independent demeaned standard deviation (EIDSD). As we have also noted, potential complications may arise from non-normality, cross-sectional dependence, or autocorrelation of the regression residuals. To circumvent those problems, the statistical significance of the impact of disputes was additionally tested using the bootstrap methodology. We shall now consider additional issues in order to test the robustness of our results. We begin with a few comments about the likelihood of additional sources of biasedness of our results. In addition, we shall deeper explore the sources of the abnormal volatility in order to provide a stronger justification for our hypothesis.

\textit{Comments on biasedness}.  Given the size of our sample, it could be argued that our result are biased due to the relatively small size of our sample. We have, therefore, taken additional steps to address the issue. We have chosen the time window during which the volatility is measured and assessed long enough in order to minimize the possibility of a small sample bias. \textit{Pari passu}, it could be argued that our sample is biased by volatility of shares of a large company that may be dominating our sample. Once again, such concerns are not justified. To recall, we have estimated the excess volatility of each stock in our sample, produced 46 excess volatilities and we report the median. We should also note that reporting the mean does not influence the main results.  We have checked the estimates of each case and have found no major deviation among them. In addition, there is no strong theoretical reason to believe that volatility of large firms should be different -- larger or smaller -- than volatility of shares of smaller companies. Finally, there is a solid econometric ground to argue that our results cannot be driven by ``excessive'' volatility of shares of single large company.  Using the GARCH model,  it is clear that an ``excessive'' volatility of shares of a large firm would affect not only our sample but also the volatility of our benchmark. 

\textit{Origins of Volatility}. Why should we observe these particular patterns of volatility? Are there any reasons beyond saying that the volatility was brought about by uncertainty about the dispute? Can we justify linking the abnormal volatility to particular features of the host countries and/or features of the arbitration process? To explore these issues, we have built a model in which we have tested the influence of factors that might have determined the link between uncertainty and volatility of share prices.  First, the size of award should clearly be important. It obviously matters whether compensations to investors are small or large in relation to damages or to claims for damages.\footnote{As noted in Section 2, the level of compensation is important particularly because it should cover not only lost profits in the past but also profits that investors could reasonably expect in future.} We expect the level of compensation to have a negative coefficient. The larger the proportion of the award from the original claim, the less uncertainty about the recovery of damages incurred by the investor and hence the “better” the impact of the award on the financial position of the investor. Second, many disputes involved host countries with several cases against them. The fact that a country is facing multiple disputes is serious problem; it tarnishes the reputation of the country and makes it more risky. In addition, if the host country in question is a poor country, the exposure to disputes clearly becomes a financial problem for the government. Third and pari passu, the uncertainty is likely to be affected by political instability in host country \citep{kurz2005determinants,irshad2017relationship}. For example, the government claiming that the conditions of the BIT were agreed and signed by the previous government may not be acceptable and may  not accept the decision of the court in favor of investor. Fourth, according some international legal scholars, the location of arbitration and/or the choice of arbitral rules may also matter \citep{pohl2012dispute,howselevin2020}.  Fifth, volatility may also be affected by the country of origin of investor. As it is well known from the financial literature, the appetite for risk is to a large extent determined by culture (e.g. \cite{li2013does,kreiser2010cultural,mihet2013effects}). Finally, a very large proportion of disputes between states and investors involve protection of public interest (e.g. \cite{schill2018wherefore,johnson2015investor}). Governments have sometimes taken measures to protect domestic public interest, and those measures have led to legal challenges by foreign investors. The range of actions (or inactions) by host governments has been large, and it would be beyond the scope of this paper to test the significance of each intervention or policy stance.\footnote{As argued by \cite{johnson2015investor}, ``Available evidence regarding the approximately 600 known ISDS suits filed to date indicate that investors can use the mechanism to contest a virtually unlimited range of actions (or inactions), including measures relating to taxation, environmental regulation, tariffs for water and electricity, health insurance regulation, and health and safety regulations of pharmaceutical imports, among others.''} As a first step, we shall try to capture the role of public interest in the litigation of investment disputes by looking at the role of two most contested rules from IIAs -- indirect expropriation and ``full protection and security'' of investors. Both represent two areas in which public interest and policy are said to have been at the origin of breaches of commitments under those agreements. The lack of clarity of both provisions is undoubtedly an important origin of uncertainty in the markets. Unfortunately, there is a strong reason to believe that, the two factors are also strongly correlated,\footnote{Many cases of ``indirect expropriation'' were also classified as breaches of ``full protection and security'', presumably reflecting plaintiffs' charges as violations of both provisions. We tried but were unable to appropriate econometric adjustments to deal with this correlation problem. } and we had to limit our analysis to ``indirect expropriation'' only.  Moreover, the unpredictability of decisions under these two provisions is further enhanced by disagreements among economists about the impact of regulatory measures on investment.\footnote{Economists disagree about whether police-powers carve-outs promote foreign investments or not. ``Weak'' policy carve-outs may promote FDI and ,with guaranteed compensation, may lead to overinvestment. Strong policy carve-out may lead to underinvestment. However, \cite{aisbett2010compensation} argues that ``underinvestment'' may happen even with guaranteed compensation. The debate has led to the literature on optimal investment agreements. See, for example, \cite{horn2019economics}.} Our model takes the following form:
\begin{equation}
\mathcal{AV}_i = \alpha + \beta_{\text{SA}} \text{SA}_i + \beta_{\text{PIC}} \text{PIC}_i + \beta_{\text{RL}} \text{RL}_i + \beta_{\text{PIC},\text{RL}} \text{PIC}_i \times \text{RL}_i + \beta_{\text{AR}} \text{AR}_i + \beta_{\text{CO}} \text{CO}_i + \beta_{\text{IE}} \text{IE}_i + \epsilon_i
\end{equation}
Where $\mathcal{AV}_t$ is abnormal volatility calculated as the volatility ratio  constructed by dividing the variance of real returns  over the event window  by the variance of real returns in the pre-event window. Our independent variables are $\text{SA}$ for the size of the award, $\text{PIC}$ for political instability and corruption, $\text{RL}$ for rule of law, $\text{AR}$ for the type of arbitral rules, $\text{CO}$ for country of origin, $\text{IE}$ for ``indirect expropriation'' as the reason for the breach of investor's rights. On the other hand, we also expect a positive effect on volatility in the case of  the multiple dispute country and political instability variables.

Unfortunately, we were also constrained by the size of our sample. As noted above, our original sample included 46 observations but as we had to restrict our analysis to cases actually won by investors, we have ended up  with 29 observations. 17 observations were ``lost'' since they referred to disputes settled out of court or to the disputes that were abandoned. Suspecting some degree of correlation, we have included in the model an interaction term between the variables captured by $\beta_{\text{PIC},\text{RL}}$ , which turns out insignificant but it leaves all other estimates fundamentally unchanged. The terms provides, therefore, an important statistiacal control. 

\begin{table} [ht!]
  \centering
  \caption{Explanation of excess volatility}   \label{tab:reg}
  \begin{threeparttable}
  \footnotesize{
  \centering
  \begin{tabular}{rrrrrl}
  \toprule
       &Estimate &Std. Error &t value &P-val & \\    
       \cmidrule{2-5}
Intercept  					& 1.044  & 0.174 & 6.004 &0.000 &  *** \\
$\beta_{\text{SA}}$         & -0.073 &  0.040& -1.813 & 0.082 & *    \\
$\beta_{\text{RL}}$         & -0.365 &  0.166& -2.193 & 0.038 &**   \\
$\beta_{\text{PIC}}$        &  0.416 &  0.199&  2.090 & 0.047 &**   \\
$\beta_{\text{AR}}$        & -0.184 &  0.105& -1.747 & 0.093 &*   \\
$\beta_{\text{CO}}$        & -0.214 &  0.118& -1.813 & 0.082 &*   \\
$\beta_{\text{IE}}$        &  0.256 &  0.139&  1.847 & 0.077 &*   \\
$\beta_{\text{PIC},\text{RL}}$      &  0.057 &  0.050&  1.140 & 0.265 &  \\ 
\cmidrule{1-5}
$adj R^2$ & & & & 0.3281 \\
\bottomrule
  \end{tabular}
    \caption*{\scriptsize \textit{Notes}: The table presents results of regression linking the abnormal volatility induced by investment disputes to independent varianbles $\text{SA}$ -- size of the award, $\text{PIC}$ -- political instability and corruption, $\text{RL}$ -- rule of law, $\text{AR}$ -- the type of arbitral rules, $\text{CO}$ -- country of origin, $\text{IE}$ -- ``indirect expropriation'' as the reason for the breach of investor's rights.. The dependent variable is defined as natural logarithm of the volatility ratio constructed by dividing the return variance computed over the event window by the variance of returns in pre-event window.}}
  \end{threeparttable}
\end{table}

As we can see from Table \ref{tab:reg}, we can explain a great deal of the abnormal volatility in our sample of foreign investment disputes. The level of compensation as a proportion of the investor's claim had a negative impact on the change in volatility even if the estimate is only significant at the 5 percent level. Both corruption and rule of law are significant and with the right sign. As expected, the results are also affected by the country of origin of investors. The results are sensitive as to whether investors come from the US or not. Interestingly enough, the results are also sensitive to two particular features of the arbitration system -- (i) whether the legal proceedings are handled under UNCITRAL or not and (ii) whether the subject of the alleged breach is ``indirect expropriation'' or not. Surprisingly, length of the process turned out to be insignificant. The adjusted R-square explains 32\% of the abnormal volatility.

The greater sensitivity of the market to dispute proceedings handled under UNCITRAL may have different explanations. For example, it is widely accepted that UNCITRAL arbitral tribunals have historically been more permissive with respect to allowing the claims of dual nationals. UNCITRAL tribunals are also less likely to refuse jurisdiction on the basis that a dispute arises from an economic activity that does not qualify as an investment. UNCITRAL awards may face additional scrutiny at the level of State courts, which may be either positive or negative.\footnote{Choosing ICSID or UNCITRAL Arbitration for Investor-State Disputes,  29/06/2018 by Aceris Law LLC \url{https://www.acerislaw.com}} Disputes about host country interventions that are considered by investors as measures leading to ``indirect expropriation'' are clearly a highly destabilizing factor in the market. The main source of unpredictability of disputes about the rule is uncertainty about the interpretation by the host country government of what the government considers the legitimate police-powers carve-out.

\section{Conclusion}

Our main aim in this paper is to assess the impact of international investment disputes on the value of the investors' firm. We do so by analyzing the second movement of the impact on shares -- on their volatility. Using the GARCH model, we estimate the cumulative average volatility of shares in our sample and compare our estimates against the benchmark. The key aspect of the methodology is to capture the usual time varying volatility, which allows us to measure abnormal variation in response to an event. In contrast to standard event studies, we do not use a single day window but a window of few days. 

We find that investment disputes lead to abnormal volatility of shares of companies involved in disputes with host countries. Markets do respond to information about the emergence of disputes between those companies and host countries. We associate the increased volatility with increased uncertainty in the market about the outcome of arbitration which may or may not lead to compensation for damages claimed by the investor. Our second important result is that the outcome of arbitration is also very important in explaining the volatility of shares of both ``winning'' and  ``losing'' investors. When the decision from arbitration went in favor of state, volatility of shares of the investors clearly increased. In contrast, when the decision of arbiters went in favor of investors, volatility of their shares visibly declined. We explain this diverging pattern by the likely impact of the decisions on the investors' balance sheets and hence the value of their firms. The impact is negative in the former case and positive in the latter. 

In order to see whether our results and the story explaining them are plausible and believable, we have conducted a number additional tests of robustness. As noted, the risk of small sample bias was minimized by applying the bootstrap method. Abnormal volatility is further controlled for stock market volatility.  We also take measures to ensure that firm size is not influencing our estimates of factors of volatility. Most importantly, to test the significance of the results, we compute the cumulative abnormal volatility statistics for different  windows of 2 days, 1 week, 2weeks, 1month and 2 months before as well as after the outcome day.  

Given the fact that there is a strong evidence of volatility, the key question is how uncertainty was affected by factors related to the arbitration awards or to the specific features of the markets. We have identified several variables that we believe should play a role in the risk assessment such as size of the award, the length of the arbitration process, the number of disputes in which host country is involved. In addition, we included four factors that have been identified in the literature -- political instability, location of arbitration, country of origin of investor and public policy considerations in host country. Using a regression analysis we find that all these factors, with the exception of the length of process are statistically significant and explain a large part of the abnormal volatility linked to the disputes in point. 

\bibliography{disputes}
\bibliographystyle{chicago}

\end{document}